\documentclass[12pt]{article}
\usepackage{amsfonts}
\usepackage[dvips]{graphicx}
\usepackage[bookmarksnumbered=true,breaklinks=true]{hyperref}
\usepackage{parskip}
\newcommand{\intx}{{\dint} d^4x\,}

\newcommand{\ri}{{\rm i}}

\newcommand{\bc}{{\bar c}}
\newcommand{\hd}{{\hat d}}
\newcommand{\ethh}[1]{\frac{\e_{#1}}{\th}}

\newcommand{\tri}{\D}
\newcommand{\var}{\dfud}
\newcommand{\Act}{{S}}
\newcommand{\inv}[1]{\frac{1}{#1}}
\newcommand{\St}{S_{\rm tot}}
\newcommand{\bS}{{\bar S}}

\newcommand{\hAstar}{{{\hat A}^*{}}}
\newcommand{\xid}{\xi^i\pa_i}

\newcommand{\tk}{{\tilde k}}

\def\ftoday{{\sl {Le \number\day \space\ifcase\month 
\or janvier\or f\'evrier\or mars\or avril\or mai
\or juin\or juillet\or ao\^ut\or septembre\or octobre
\or novembre \or d\'ecembre\fi\space \number\year}}}    
\def\ptoday{{\sl {\number\day \space de\space \ifcase\month 
\or janeiro\or fevereiro\or mar{\c c}o\or abril\or maio
\or junho\or julho\or agosto\or setembro\or outubro
\or novembro \or dezembro\fi\space de\space \number\year}}}    
\def\gtoday{{\sl {Den \number\day. \ifcase\month 
\or Januar\or Februar\or M\"arz\or April\or Mai
\or Juni\or Juli\or August\or September\or Oktober
\or November \or Dezember\fi\space \number\year}}}    
\def\today{{\sl {\ifcase\month
\or January\or February\or March\or April\or May
\or June\or July\or August\or September\or October
\or November \or December\fi \space\number\day,\space 
                                            \number\year}}}
\newcommand{\journal}[4]{\emph{#1~}{\bf #2}\,(#3)\,#4}

\newcommand{\physu}{\journal {Physica (Utrecht)}}
\newcommand{\pr}{\journal {Phys. Rev.}}

\newcommand{\rmp}{\journal {Rev. Mod. Phys.}}

\newcommand{\prep}{\journal {Phys. Rep.}}

           \newcommand{\GA}{\Gamma}
\renewcommand{\d}{\delta}         \newcommand{\D}{\Delta}
\newcommand{\e}{\varepsilon}

\newcommand{\la}{\lambda}        
\newcommand{\m}{\mu}
\newcommand{\n}{\nu}

\renewcommand{\th}{\theta}         
           \newcommand{\F}{{\Phi}}
\newcommand{\vf}{{\varphi}}

\newcommand{\SSS}{{\cal S}}

\newcommand{\es}{\\[3mm]}
\newcommand{\ES}{\\[6mm]}

\newcommand{\sla}{\raise.15ex\hbox{$/$}\kern -.57em} 
\newcommand{\Sla}{\raise.15ex\hbox{$/$}\kern -.70em}

\def\Lp{\displaystyle{\biggl(}}
\def\Rp{\displaystyle{\biggr)}}
\def\LP{\displaystyle{\Biggl(}}
\def\RP{\displaystyle{\Biggr)}}
\newcommand{\lp}{\left(}\newcommand{\rp}{\right)}

\newcommand{\complex}{{\kern .1em {\raise .47ex
\hbox {$\scriptscriptstyle |$}}
    \kern -.4em {\rm C}}}
\newcommand{\real}{{{\rm I} \kern -.19em {\rm R}}}
\newcommand{\rational}{{\kern .1em {\raise .47ex
\hbox{$\scripscriptstyle |$}}
    \kern -.35em {\rm Q}}}
\renewcommand{\natural}{{\vrule height 1.6ex width
.05em depth 0ex \kern -.35em {\rm N}}}

\newcommand{\pa}{\partial}

\newcommand{\fud}[2]{{\frac{\delta #1}{\delta #2}}}
\newcommand{\dpad}[2]{{\displaystyle{\frac{\partial #1}{\partial #2}}}}
\newcommand{\dfud}[2]{{\displaystyle{\frac{\delta #1}{\delta #2}}}}
\newcommand{\dfrac}[2]{{\displaystyle{\frac{#1}{#2}}}}
\newcommand{\dsum}[2]{\displaystyle{\sum_{#1}^{#2}}}   
\newcommand{\dint}{\displaystyle{\int}}

\newcommand{\twiddle}{\lower.9ex\rlap{$\kern -.1em\scriptstyle\sim$}}

\newcommand{\equ}[1]{(\ref{#1})}
\newcommand{\eq}{\begin{equation}}
\newcommand{\eqn}[1]{\label{#1}\end{equation}}
\newcommand{\eea}{\end{eqnarray}}
\newcommand{\eqa}{\begin{eqnarray}}
\newcommand{\eqan}[1]{\label{#1}\end{eqnarray}}
\newcommand{\ba}{\begin{array}}
\newcommand{\ea}{\end{array}}
\newcommand{\eqac}{\begin{equation}\begin{array}{rcl}}
\newcommand{\eqacn}[1]{\end{array}\label{#1}\end{equation}}

\begin{document}
\title{\bf A Vector Supersymmetry Killing the Infrared Singularity of Gauge 
Theories in 
Noncommutative 
Space }

\author{\\[-0.3cm]\Large 
Daniel N.~Blaschke\footnote{work supported by ``Fonds zur F\"orderung der 
Wissenschaftlichen Forschung'' (FWF) under contract P15015-N08.}~, 
Fran\c cois Gieres\footnotemark[3]~, 
Olivier Piguet\footnote{work supported  
in part by the Conselho Nacional 
de Desenvolvimento Cient\'{\i}fico e  
Tecnol\'{o}gico CNPq -- Brazil.} \\[1.5mm] \Large
and  Manfred Schweda\footnotemark[1]
}
\date{}

\maketitle
\begin{center}
\renewcommand{\thefootnote}{\fnsymbol{footnote}}
\vspace{-0.3cm}\footnotemark[1]Institute for Theoretical Physics, 
Vienna University of Technology\\
Wiedner Hauptstrasse 8-10, A-1040 Vienna (Austria)\\[0.3cm]
\footnotemark[3]Institut de Physique Nucl\'eaire,
Universit\'e Claude Bernard (Lyon 1), \\
4 rue Enrico Fermi, F - 69622 - Villeurbanne (France)\\[0.3cm]
\footnotemark[2]Departamento de F\'{\i}sica, CCE,
Universidade Federal do Esp\'{\i}rito Santo (UFES), \\
Av. Fernando Ferrari, 514, BR-29075-910 - Vit\'oria - ES (Brasil)\\[0.5cm]
\ttfamily{E-mail: blaschke@hep.itp.tuwien.ac.at, gieres@ipnl.in2p3.fr, opiguet@yahoo.com, mschweda@tph.tuwien.ac.at}
\vspace{0.5cm}
\end{center}

\abstract{We show that the "topological BF-type" term introduced by Slavnov in
order to cure the infrared divergences of gauge theories in
noncommutative space 
can be characterized as the consequence of 
a new symmetry. This symmetry is a supersymmetry, generated by vector charges,
of the same type as the one encountered in Chern-Simons or BF topological
theories.}
\vspace{5mm}

{\it Work presented by O. Piguet at the 
Fifth International Conference on Mathematical Methods 
in Physics, 24 - 28 April 2006, Rio de Janeiro, Brazil}

\section{Introduction}

The idea of noncommuting position and time coordinates~\cite{review}
 has been first 
introduced in the literature in Ref. ~\cite{Snyder}. 
In flat space-time, this amounts to postulate that the 
coordinates
obey commutation relations
\eq
[x^\m,x^\n]= \ri \th^{\m\n}\,,
\eqn{noncomm}
for some antisymmetric matrix $\th^{\m\n} = -\th^{\n\m}$. Classical
fields $\vf(x)$ thus become noncommuting because of the noncommutativity of 
the $x^{\mu}$. 
An equivalent implementation of noncommutativity is 
given by 
the Moyal~\cite{Moyal} product, which is associative, but noncommutative:
\eq
\vf_1(x)*\vf_2(x) =\left. e^{\dfrac{\ri}2\th^{\m\n}\pa_\m^x\pa_\n^y}
\lp \vf_1(x)\vf_2(y) \rp\right|_{y=x}\,,
\eqn{moyal-product}
where the coordinates are considered as commuting\footnote{From now on,
all field products will be Moyal ones, and the
symbol $*$ will be omitted.}. We consider a 
(1+3)-dimensional  
space-time
with Minkowski metric  $(\eta_{\m\n})=$ diag $(1,-1,-1,-1)$.

The field theory under consideration contains a U(1) gauge field $A_\m$ 
and a scalar field  $\la$ (Slavnov's field), with infinitesimal 
gauge transformations
\eq
\d A_\m= \pa_\m\e - \ri g [A_\m,\e]\,,\quad \d\la= -\ri g [\la,\e]\,.
\eqn{gauge-tr}
Notice the presence of commutators, $[X,Y]$ = $X*Y-Y*X$, due to the
noncommutativity of the Moyal product.

\begin{figure}[h]
\centering
\includegraphics[scale=0.6]{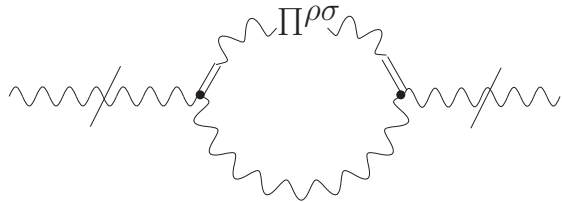}
\caption{IR divergent graph with vacuum polarization insertion.}
\label{fig1}
\end{figure}

It is well known~\cite{Filk,Matusis,Blaschke1,review} 
that such gauge theories in
noncommutative space-time -- which 
are 
renormalizable 
in commutative space-time --  suffer from infrared (IR) singularities mixed
with the usual ultraviolet (UV) divergences. 
Indeed, a gauge invariant action such 
as\footnote{Remember that all products are Moyal.}
\eq
S_{\rm Maxwell} = -\dfrac{1}{4}\intx \  F_{\mu\nu} F^{\mu\nu} \,,
\eqn{maxwell-action}
possibly coupled with matter fields, with 
\eq
F_{\m\n}=\pa_\m A_\n - \pa_\n A_\m -\ri g [A_\m,A_\n]\,,
\eqn{field-strength}
 leads to infrared (IR) singularities
associated with ultraviolet (UV) divergent Feynman diagrams. Typically,
vacuum polarization graphs have IR singular parts
\eq
\Pi^{\mu\nu}_{\rm{IR}}(k)
=\dfrac{2g^2}{\pi^2} \, \dfrac{\tk^\mu\tk^\nu}{(\tk^2)^2}\,,\quad
\mbox{with}\ \tk^\m = \th^{\m\n}k_\n\,,
\eqn{vac-pol}
and graphs with this insertion,  such as the one shown in Fig.~\ref{fig1}, 
are IR divergent.

Only special gauge theories are known to be free from these divergences 
(see e.g. the review~\cite{Gomes}). Among them, let us mention 
Chern-Simons topological theory with particular 
couplings to matter~\cite{Bichl}, BF theories~\cite{Blasi}
and some supersymmetric Yang-Mills 
theories~\cite{Asano}. These theories are or topological~\cite{Birmingham}, 
or supersymmetric~\cite{Wess}. 
A relevant question 
to ask is whether 
 nontopological gauge theories 
on noncommutative space 
need to be
supersymmetric in order to be free of IR singularities. A partial 
answer to this question is the object of the rest of this talk, which
summarizes results presented in Ref.~\cite{Blaschke3}

\section{Slavnov's modification of the noncommutative U(1) theory}

Slavnov~\cite{Slavnov1,Slavnov2} has proposed a modification of the theory, 
adding to the action 
\equ{maxwell-action}
the term
\eq
\dfrac{1}{2}
\intx \, \la\, \th^{\mu\nu}F_{\mu\nu}
\,,
\eqn{slavnov-term}
which involves the scalar field $\la$ as a Lagrange multiplier. This term
looks like a topological $BF$ action. It reduces the 
degrees of freedom of a 
spin 1 gauge boson
to those 
of a spin 0 particle -- whereas 
the suppression of the local degrees of freedom is complete
in a true topological theory.
Slavnov has shown 
by 
a power-counting argument 
that IR singularities are absent in the theory obtained by adding the term
\equ{slavnov-term} to the Maxwell action \equ{maxwell-action} -- 
see also~\cite{Blaschke2}. We shall
point out that the absence of IR 
singularities
is in fact a consequence of 
the invariance of the theory under a
vector supersymmetry.

In the following we shall choose the noncommutativity tensor to 
be space-like,
in order to avoid problems with unitarity hence, without loss of generality, 
in the (1,2)--plane:
\[
\th^{ij}= \th\e^{ij}\,,\quad i,j = 1,2  \,\quad(\th^{12}=-\th^{21}=1)\,.
\]
We use the notation $i,j,\cdots=1,2$ and $I,J,\cdots = 0,3$.

The gauge invariant action thus reads
\eq
S_{{\rm inv}}[A,\la] =
\intx  \lp -\dfrac{1}{4}F_{\mu\nu}(x) F^{\mu\nu}(x)
+ 
\dfrac{1}{2} \la(x)  \, \th^{ij}F_{ij}(x) \rp
\eqn{g-inv-action}
with $F_{\mu\nu}$ given by \equ{field-strength}.

\section{Gauge fixing and BRST symmetry}

Gauge fixing will conveniently be chosen axial, 
in the plane of the noncommutative
coordinates. It is characterized by a vector 
$
(n^{\mu})
=(0,1,0,0)$, a Lagrange
multiplier field $B(x)$ and Faddeev-Popov ghosts 
$\bc(x)$, $c(x)$. The complete action is
\eq\ba{c}
S= S_{{\rm inv}}[A,\la] + S_{{\rm gf}}[A,B,c,\bc]\,,\es
\mbox{with} \quad 
 S_{{\rm gf}}[A,B,c,\bc] =
 \intx \lp B(x) n^iA_i(x)-\bc(x) n^iD_ic(x) \rp\, , 
\ea\eqn{g-fixed-action}
where
\[
D_{\mu} c = \partial_{\mu} c - \ri g [A_{\mu},{c}] \, .
\]
This action is invariant under the BRST transformations
\eq\ba{ll}
sA_\mu = D_\mu c \, , \qquad  &s \bar c  = B \, , \es
 s\la = - \ri g \, [\la, c ] \, , \qquad & sB = 0 \, ,  \es
sc  = \dfrac{\ri g}{2}  \, [c , c ] \, , &
\ea\eqn{BRST}
the BRST operator $s$ being nilpotent: $\ s^2=0$.
All commutators are assumed to be graded with respect to the 
ghost-number.

\section{Vector supersymmetry, superalgebra and generalized BRST operator}

We note that the Slavnov term, together with the gauge-fixing terms, have
the form of a 2-dimensional gauge fixed topological $BF$ model, $\la$ 
playing the role of the ``$B$'' field. Topological
models of this kind (``Schwarz type'' topological models) 
are known to possess a symmetry generated by a fermionic vector
charge, called vector supersymmetry (VSUSY),
responsible for their UV finiteness~\cite{Delduc,Blasi-P-S,Emery}.
It turns out that this is 
also true
here: the total gauge-fixed action 
is invariant under the {VSUSY transformations}
\eq\ba{lll}
\d_iA_J =0\,, \quad&\d_iA_j = 0 \, ,  \quad 
   &\d_i\la = \dfrac{ \e_{ij} }{\th} n^j\bc \, ,\es
\d_ic=A_i \, ,\quad &\d_i\bc=0 \, ,\quad 
   &\d_iB=\partial_i\bc \, .
\ea\quad(i=1,2)
\eqn{VSUSY}
In the pure topological theories, the BRST and VSUSY generators form a
closed algebra together with the translation generators. Here, in order
to have a closed algebra, we must invoke an additional 
vector symmetry of the
gauge-fixed action, peculiar to the present theory:
\eq\ba{l}
\hd_i A_J =  -F_{iJ}\,,
\quad \hd_i\la = -\dfrac{\e_{ij}}{\th}
D_K F^{Kj}
\,, \es
\hd_i\Phi =0  \quad 
\mbox{for all other fields} \, .
\ea \quad (i=1,2)
\eqn{add-SUSY}
The  algebra involving $s$, $\d_i$,  $\hd_i$ and the (1-2)-plane translation
generators $\pa_i$ is closed -- modulo equations of motion:
\eq
\left.\begin{array}{l}
[\pa_i,s]\Phi = [\pa_i,\d_j]\Phi =[\pa_i,\hd_j]\Phi = 0\,,  \es
 {[s, s]} \, \Phi = 
{[s, \hd_j]} \, \Phi = 0\,, \es
{[ \d_i, \d_j ]} \, \Phi
= {[ \d_i , \hd_j ]} \, \Phi = 0\,,
\end{array} \right\} 
\quad  \mbox{for all fields} \ \Phi \,,  \es
\eqn{s-algebra-s}
\eq\ba{l}
{[s,\d_i]}\, \Phi =\partial_i  \Phi + \hd_i \Phi  
\qquad \quad {\rm for} \ \Phi \in \{ A_J, c, \bc, B \} 
\,, \label{algebra2'}\es
{[s , \d_i]} \, A_j = \partial_i A_j+ \hd_i A_j
-\dfrac{\e_{ij}}{\th} \dfud{ S}{\la}\, ,\es
{[s, \d_i]} \, \la =\partial_i \la+ \hd_i\la+\dfrac{\e_{ij}}{\th}\dfud{ S}{A_j}
- \dfrac{1}{\th^2} D_i\dfud{ S}{\la} \,,\es
\ea\eqn{s-algebra-vsusy}
\eq\ba{l}[ \hd_i , \hd_j] \, A_J = 
\dfrac{\e_{ij}}{\th}D_J\dfud{ S}{\la}   \,,  \es
{[\hd_i, \hd_j ]} \, \la =
\dfrac{\e_{ij}}{\th}D_J\dfud{ S}{A_J}\,,   \es
{[ \hd_i , \hd_j]} \, \Phi = 0
\qquad \qquad \quad {\rm for} \ \Phi \in \{ A_i, c, \bc, B \} 
\, . 
\ea\eqn{s-algebra-add-susy}

The various symmetries will be combined into a
single generalized BRST operator $\D$, with the 
introduction of the constant ghosts
$\xi^i,\, \e^i,\, \m^i$ playing the role of the infinitesimal parameters
of the symmetries  $\pa_i,\, \d_i,\, \hd_i$. The statistics of these 
constant ghosts is fermionic, bosonic and fermionic, respectively. The
generalized BRST operator thus reads
\eq
\D = s + \xi^i\pa_i + \e ^i \d_i+\mu^i\hd_i\,,
\eqn{gen-BRST}
and its action on the various fields and on the constant ghosts is given by
\eq\ba{l}
\D A_i=D_ic
+\xi^j \pa_j  A_i
\, ,\es
\D A_J=D_Jc+\xi^i\pa_i  A_J+\mu^iF_{Ji}
\, ,\es
\D\la=- \ri  g \, {[ \la, c]} + \xi^i\pa_i  \la+\e^i \ethh{ij}n^j\bc
+\mu^i\ethh{ij}D_KF^{jK}
\, ,\es
\D c= 
\dfrac{\ri g}{2} \, {[c,  c] } 
+\xi^i\pa_i  c+\e^i A_i
\, ,\es
\D\bc=B+\xi^i\pa_i  \bc
\, ,\es
\D B=\xi^i\pa_i  B+\e^i\pa_i\bc
\, , \es
\D\xi^i= \D \mu^i=-\e^i \, , \quad  \D\e^i=0 \, .
\ea\eqn{Delta-fields}
$\D$ is nilpotent, but only on-shell:
\eq\ba{l}
\tri^2 A_i=\e^j\ethh{ij}\var{\Act}{\la} 
\, ,\es
\D^2 A_J=\dfrac{\mu^i\mu^j}{2}\ethh{ij}D_J\var{\Act}{\la}
\, ,\label{nilpoten-op2}\es
\D^2\la=\dfrac{\mu^i\mu^j}{2}\ethh{ij}D_J\var{\Act}{A_J}
+\e^i\ethh{ij}\var{\Act}{A_j}-\e^i\inv{\th^2}D_i\var{\Act}{\la} \, ,\es
\D^2 c = \tri^2 \bc = \tri^2 B = 0 \, .
\ea\eqn{Delta-nilpotent}

\section{Slavnov identity and ghost equations}

Useful Ward identities are consequences of the
Slavnov-Taylor identity
describing the invariance of the theory under the transformations
\equ{Delta-fields}. In order to write 
this identity, 
we associate an external field $\F^*$ --
an ``antifield'' in the terminology of the authors of Ref.~\cite{Batalin} -- 
to the  $\D$-variation of each of the fields $\F$ = $A,\,\la,\,c$, 
respectively. The action $\St$ 
depending on the fields and antifields is a solution of 
the Slavnov-Taylor identity
\eq\ba{l}
\SSS(\St)\equiv 
\dint d^4x \, 
\Lp \dsum{\Phi \in \{ A_{\mu} ,\la,c \}}
{}
\dfud{\St}{\Phi^*}\dfud{\St}{\Phi}
\, + \, \lp B+ \xi^i\pa_i \bc \rp \dfud{\St}{\bc} 
 \quad
+ \, \lp \xi^i\pa_i B + \e^i\pa_i \bc \rp
\dfud{\St}{B} \Rp
\es\phantom{\SSS(\St)\equiv }
-\e^i ( \dpad{\St}{\xi^i} + \dpad{\St}{\mu^i} ) =0 \, . 
\ea\eqn{Slavnov-id}
The solution reads
\eq\ba{l}
\St[ A,\la,c,\bc, B \, ;A^*,\la^*,
c^* ;\xi,\mu,\e] \es\qquad
= \dint d^4x\, (B + \xi^i\pa_i \bc ) n^i A_i 
 +  \bar S[A,\la,c \, ; \hAstar^i, A^{*J}  ,\la^*,c^*; \, \xi,\mu,\e]
\, ,
\ea\eqn{sol-slavnov-id}
where $\hAstar^i = A^{*i}  -n^i\bc$, and
\eq\ba{l}
\bS =
\dint d^4x\, \LP -\dfrac{1}{4}F_{\mu\nu} F^{\mu\nu}
+ \dfrac{\th}{2} \la \e^{ij}F_{ij}  \es\qquad
+ \, { {\hAstar^i}} \left( D_ic + 
\xi^{j} \partial_j 
A_i \right)
+{ {A^*}^{J}} \left( D_Jc+\xid A_J+\mu^iF_{Ji} \right)  \es\qquad
+ \, { {\la^*}} 
\left( -\ri g {[\la ,c] } +\xid \la
+\mu^i \dfrac{\e_{ij}}{\th} D_KF^{jK} \right)
+{ {c^*}} \left( 
\dfrac{\ri g}{2} \, {[c,c]}+\xid c+\e^i A_i \right)  \es\qquad
+ \left( \dfrac{\mu^i\mu^j}{2}\, \dfrac{\epsilon_{ij}}{\th} 
(D_J { {A^*}^{J}} )
+ \e^i \dfrac{\epsilon_{ij}}{\th} \, { \hAstar^j}
- \e^i 
\dfrac{1}{2\th^2} \, (D_i{{ \la^*}} ) \right) { {\la^*}} 
\RP
\, .
\ea\eqn{trunc-action}

Due to the axial gauge fixing, the field equations for $c$ and $\bc$ take
the form of local functional equations, namely, the antighost equation:
\eq
\dfud{\St}{c} + \ri g\left[\bc,\dfud{\St}{B}\right] \ES
\quad = - n^i\pa_i\bc +
D_\mu{A^*}^\mu -\ri g[\la,{\la^*}] + \ri g[c,{c^*}]
+\xid{c^*}\,, 
\eqn{antighost-eq}
and the ghost equation:
\eq
\dfud{\St}{\bc} + \ri g\left[c,\dfud{\St}{B}\right] 
-\xid \dfud{\St}{B} \ES
\quad = -  n^i\pa_i c - \e^i \dfrac{\epsilon_{ij}}{\th}n^j{\la^*}\,.
\eqn{ghost-eq}
Note that both right hand sides are linear in the quantum fields. This
fact expresses the well-known freedom of the ghost fields  in axial
gauges~\cite{Kummer}.

\section{Ward identities of vector supersymmetry}

Interesting Ward identities may be extracted from the 
Slavnov-Taylor identity and from the ghost and
antighost equations. E.g., a  Ward identity for VSUSY
is obtained by differentiating the Slavnov-Taylor identity 
\equ{Slavnov-id} with respect to the VSUSY ghost $\e^i$. The result is
\eq
{\cal W}_i \St= \Delta_i 
\, ,
\eqn{vsusy-WI}
with
\[\ba{l}
{\cal W}_i \St= \intx\, \LP\partial_i\bc \, \var{\St}{B}
+A_i \, \var{\St}{c} 
+\left( \ethh{ij}\left(n^j\bc-{A^*}^j\right)
+\dfrac{1}{\th^2}
D_i{\la^*}\right)\var{\St}{\la} \es\phantom{{\cal W}_i \St= \intx\, \LP}
 + {\la^*} \ethh{ij} \, \var{\St}{A_j}
+\left({c^*}+ \dfrac{\ri g}{\th^2}{\la^*}{\la^*}\right)\var{\St}{{A^*}^i}\RP\,,
\ea\]
and 
\[
\Delta_i =
\dpad{\St}{\xi^i}+\dpad{\St}{\mu^i}
+ \intx\, 
\ethh{ij}n^j\left(B+\xid \bc\right){\la^*}\, .
\]
We note that the breaking term $\D_i$ vanishes at vanishing antifields.


Knowing that the total action 
$
\St
[\F,\F^*,\cdots]$ is
the functional generator of the vertex
functions (1-particle irreducible amputated graph contributions) in the
tree graph approximation, and that the Legendre transform
\[
Z^{\rm c}[J_\F,\F^*,\cdots] = \St[\F, \F^*,\cdots] + 
\sum_\F\int 
d^4x \, 
J_\F\F\,,\quad \mbox{with}\quad J_\F = -\fud{\St}{\F}\,,
\] 
yields the functional generator of the connected Green functions, 
we obtain Ward identities for the connected 
Green functions -- here in the tree approximation. The Ward identity 
for VSUSY at vanishing antifields, which reads, for the vertex
functions, as
\eq
\intx\, \lp\partial_i\bc \, \var{\St}{B}
+A_i \, \var{\St}{c} 
+ \ethh{ij} n^j\bc \var{\St}{\la}\rp = 0\,,
\eqn{wivert}
yields, for the connected Green functions, 
\[
\int d^4x \lp 
j_B \, \partial_i \var{Z^c}{j_\bc}   
- j_c \, \var{Z^c}{j_A^i} \, 
+ \ethh{ij} n^j  j_\la \,  \var{Z^c}{j_\bc} \,  \rp=0 \, .
\]
Differentiating, 
e.g. with respect to $j_c$ and to $j_A^\n$,
yields, for the gauge field propagator, the condition
\eq
\D_{A_i A_\m}=0\,.
\eqn{WI-prop}
Other consequences of VSUSY are
obtained from the Ward identity for vertex functions 
\equ{wivert} by differentiating 
with respect to $A_\m$ and $A_J$, or $A_i$ and $A_j$:
\eq
\GA_{\la A_\m A_J}(x,y,z)=0\,, \es
\eqn{WI-vertices1}
and
\eq
\GA_{\la A_i A_j}(x,y,z)= \ri g\th\e^{kl}K(x,y,z)\,,
\eqn{WI-vertices2}
where
\[
K(x,y,z) =  e^{\dfrac{\ri}2\th\e^{ij}\pa_i^x\pa_j^u}
\left. \lp \d(x-y)-\d(u-z) - \d(x-z)-\d(u-y) \rp \right|_{u=x}\,.
\]
If we assume the latter result, which coincides with the tree vertex
deduced from the classical action $\St$, to be valid for the quantized theory,
we 
could 
conclude that the vertex
$\GA_{\la A_i A_j}$ 
does not acquire 
radiative corrections.

\section{Cancellation of the IR singularities}

\begin{figure}[h]
\centering
\includegraphics[scale=0.9]{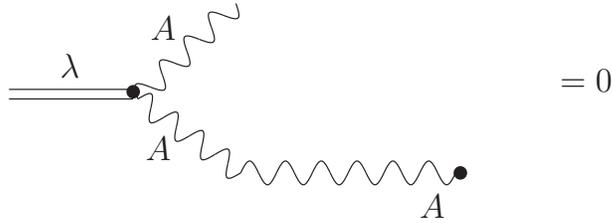}
\caption{The $\la AA$-vertex contracted with a photon propagator.}
\label{fig2}
\end{figure}

\begin{figure}[h]
\centering
\includegraphics[scale=0.9]{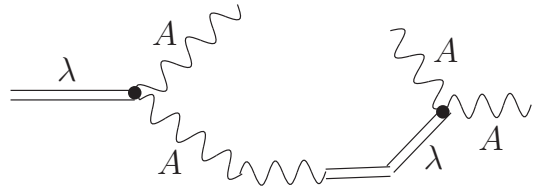}
\caption{Trying to build loop graphs with a $\la AA$-vertex 
and $A\la$-propagators,
but without $AA$-propagators.}
\label{fig3}
\end{figure}

Let us now show 
by 
a graphical analysis 
how the IR singularities are cancelled as a consequence
of the VSUSY Ward identities. We first see, from Fig.~\ref{fig2},
that the $\la AA$-vertex contracted with a $AA$-propagator vanishes
because of \equ{WI-prop}.
Secondly, looking at Fig.~\ref{fig3},
we observe that one cannot build a Feynman loop graph containing a 
$\la AA$-vertex  without the presence of at least one $AA$-propagator.
Thus, it 
follows that loop corrections to 
the $\la\la$ and $\la A$ propagators vanish.

More generally, we can conclude that
all loop graphs involving a $\la AA$ vertex vanish.
In particular, dangerous vacuum polarization insertions as in 
Fig.~\ref{fig1} cancel.
Finally, contributions of IR singular parts of vertices
$\GA_{A_{\m_1}\cdots A_{\m_N}}$  ($N\ge2$) connected to $AA$-propagators 
 in loop graphs vanish, too, since these singularities 
are present in vertices with indices $i=1,2$ only. E.g. in the 
singular part \equ{vac-pol} of the vacuum polarization, 
the indices $\m$ and $\n$
take the values 1 or 2 due to our choice of the noncommutativity matrix
$\theta$.

In conclusion, no IR singularities are left.

\section{Conclusions and 
outlook
}

We can conclude from our analysis that Poincar\'e supersymmetry 
is not needed in order to cure the 
problem of 
IR-UV mixing in gauge theories constructed in noncommutative space.

However the concept of supersymmetry, manifest in the form of VSUSY, seems to
play a decisive role in theories which are not Poincar\'e
supersymmetric. 
Indeed, as we have seen in such a case, the Ward identities 
of VSUSY yield exactly 
the propagator and vertex properties which are needed for cancelling the
IR singularities.

What is the role of VSUSY with respect to the IR-UV mixing
in topological gauge theories in general remains an open question.

Finally, the study of more general theories, based on a rigorous quantization
scheme  (perturbative~\cite{Piguet-Sorella} or not) seems desirable.



\end{document}